\documentclass[runningheads]{llncs}

\usepackage[T1]{fontenc}
\DeclareUnicodeCharacter{2113}{\ell}
\usepackage[utf8]{inputenc}
\usepackage{hyperref}
\usepackage{graphicx}
\usepackage{makecell}
\usepackage{multirow}
\usepackage{eucal}
\usepackage{mathptmx}
\usepackage{mathtools}
\begin{document}
\title{EB-GAME: A Game-Changer in ECG Heartbeat Anomaly Detection}
%
%
\author{June Young Park\inst{1}\ \and
Da Young Kim\inst{1}\ \and
Yunsoo Kim\inst{2}\ \and
Jisu Yoo\inst{1}\ \and \\
Tae Joon Kim\inst{3}}
%
\authorrunning{JY.Park et al.}
%
\institute{Graduate School of Ajou University, \\ Department of Convergence Healthcare Medicine, Suwon, Korea \and
Ajou University Medical Center, Suwon, Korea
\and
Ajou University School of Medicine, Department of Neurology, Suwon, Korea\\
\email{\{wnsdud1007, tjkim23\}@ajou.ac.kr}}
\maketitle              
\begin{abstract}
Cardiologists use electrocardiograms (ECG) for the detection of arrhythmias. However, continuous monitoring of ECG signals to detect cardiac abnormalities requires significant time and human resources. As a result, several deep learning studies have been conducted in advance for the automatic detection of arrhythmia. These models show relatively high performance in supervised learning, but are not applicable in cases with few training examples. This is because abnormal ECG data is scarce compared to normal data in most real-world clinical settings. Therefore, in this study, GAN-based anomaly detection, i.e., unsupervised learning, was employed to address the issue of data imbalance. This paper focuses on detecting abnormal signals in electrocardiograms (ECGs) using only labels from normal signals as training data. Inspired by self-supervised vision transformers, which learn by dividing images into patches, and masked auto-encoders, known for their effectiveness in patch reconstruction and solving information redundancy, we introduce the ECG Heartbeat Anomaly Detection model, EB-GAME. EB-GAME was trained and validated on the MIT-BIH Arrhythmia Dataset, where it achieved state-of-the-art performance on this benchmark.

\end{abstract}
\section{Introduction}
Electrocardiograms (ECG) is one of the most widely used non-invasive techniques for identifying cardiac abnormalities and usually carried out by specially trained cardiologists [1]. As such, the interpretation of ECG requires considerable time and extensive human resources. On top of that, misinterpretation may occur if there are abundant data to detect [2]. Hence, the deep neural network (DNN) model have been developed to assist the physicians and proven its capability of automated anomaly detection to solve the aforementioned issues [3]. In the real world, the supervised models encountered the class imbalance problem due to the low manifestations of abnormal cases, which can seriously affect the generalization ability of the model [4]. Numerous efforts have been made to tackle the problem, and generative adversarial net-work (GAN) variations successfully addressed its application toward the detection of abnormalities in several studies [5, 6]. \\
GAN based models basically generate the image from the original data and detect the abnormalities using reconstruction error by discriminating the differences be-tween the original image and the generated image. Among the variants of GAN, T Schlege et al. proposed the deep convolutional generative adversarial nets (DCGAN), which introduced a new anomaly scoring method based on mapping from image space to latent space without the need of annotations as learning targets during training [7]. Furthermore, S Akcay et al. suggested GANomaly as the conditional generative adversarial method. This method trains only on normal data and simultaneously generates high-dimensional image space and inference of latent space during the process [8].
As the method for anomaly detection continuously advanced, there has been a lot of research in the field of image reconstruction methods using ECG recently. Shin et al. developed a model to detect abnormal ECG beats using only normal ECG data and proposed the improved AnoGANs by calculating decision boundaries through repeated testing [5]. BeatGAN was developed to automatically detect anomalous beats. This beat based detection model used adversarial generated beats and was able to identify the time periods of abnormal beats [6]. Additionally, several attempts were accomplished to enhance the performance of reconstruction in ECG such as adding a new loss function or inserting a bi-directional long-short term memory (Bi-LSTM) layer in the generator [9, 10]. \\
Previous reconstruction-based anomaly detection methods have shown promising results, yet they suffer from the drawback of requiring computationally intensive preprocessing for heartbeat detection. Given the scarcity of ECG data, it is challenging to deploy models that demand all peaks of an ECG beat in real clinical settings. In response, we propose the ECG Beat Generative Adversarial networks-based Masked autoencoder architecture, abbreviated as EB-GAME, which reconstructs the normal beat of an ECG by masking some peaks of the beat
(see Fig. 1).

\begin{figure}
\includegraphics[width=\textwidth]{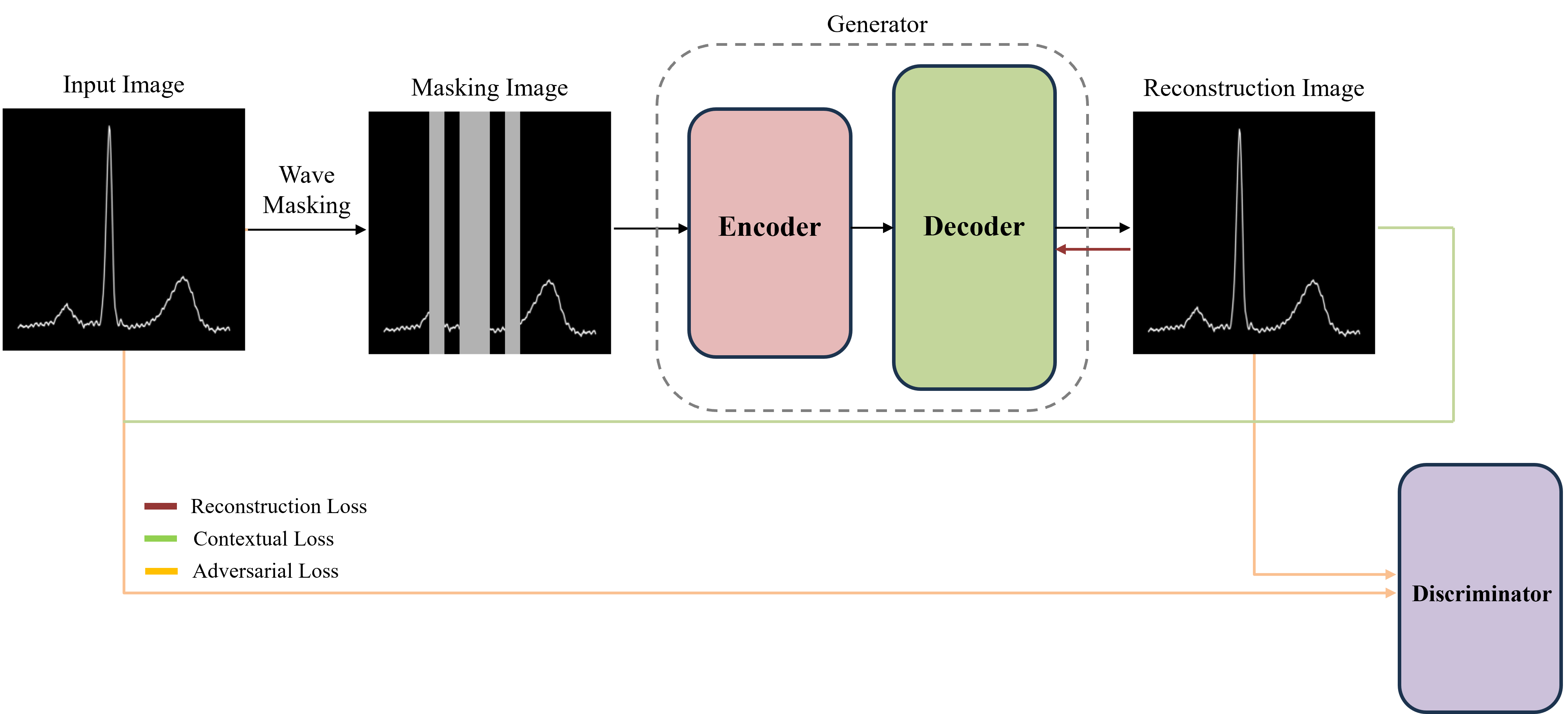}
\caption{Overview of the EB-GAME framework. The generator employs the MAE architecture, and through adversarial training with the discriminator, it is trained and performs anomaly detection.} \label{fig1}
\end{figure}

\section{Method}
\subsection{Preliminary}
\subsubsection{MAE.} Masked Autoencoder (MAE) is a form of self-supervised learning method aimed at efficiently extracting features from visual data, such as images [11]. This technique operates by randomly masking parts of the input image and then attempting to reconstruct the original image based on this masked input. The core components of MAE are divided into two parts: the encoder and the decoder. The encoder utilizes only the visible parts of the data to learn a compressed representation of the input image, while the decoder attempts to reconstruct the entire image, including the masked parts, from this latent space. The training process of the MAE primarily consists of two stages. In the first stage, random patches from an image are selected and masked, and this masked image is then fed into an encoder. The encoder extracts important information from this input and converts it into a latent representation. Following this, in the second stage, the decoder attempts to accurately restore the masked parts of the original image using this latent representation. Through this process, the model learns significant features and patterns of the image, which can be utilized in various visual tasks such as image classification and object detection. One of the main advantages of MAE is its efficient data representation learning capability. By employing a masking technique, the model can focus more on the important features of the image, thereby enhancing its understanding of the entire dataset. Moreover, MAE adopts a self-supervised learning approach, offering the benefit of utilizing large amounts of unlabeled data. This aspect is particularly useful in situations where labeling costs are high or the amount of labeled data is limited. In conclusion, the Masked Autoencoder is a powerful tool capable of efficiently extracting crucial features of visual data and leveraging this for superior performance in various visual recognition tasks. Thanks to these characteristics, MAE is widely researched in the field of computer vision and is expected to see significant advancements in the future.

\subsubsection{GAN-based Anomaly Detection.} GAN-based anomaly detection operates on an unsupervised learning approach where the generator learns the distribution of normal data, and the discriminator identifies data that deviates from this distribution as anomalies. In anomaly detection, GANs are primarily trained using only normal data, and the trained model is then used to evaluate how well new data fits the normal distribution. The further the data deviates from the normal distribution, the more it is considered an anomaly. Moreover, GAN-based anomaly detection offers a flexible solution to address various real-world problems such as data imbalance, the high cost of data labeling, and the uncertainty of labels through unsupervised learning.

\subsection{Let's GAME Change}
In this section, we introduce the EB-GAME framework, which employs a novel strategy for masking waves in ECG data and utilizes the MAE architecture to construct both the Generator and Discriminator. The cornerstone of our approach is the innovative use of wave masking, enabling more focused analysis of critical ECG segments and thereby enhancing the model's capability to identify deviations from normal patterns. By integrating this strategy with the strengths of the MAE architecture, the Generator and Discriminator are optimized for high-fidelity generation and precise anomaly detection, respectively. Finally, we define the learning objectives that are specifically designed to enable the proposed framework to effectively detect anomalies in ECG data.

\subsubsection{Wave Masking Strategy.} The standard MAE follows an encoder-decoder architecture, through which it performs the task of reconstructing images that have been processed with masking. In this process, the input image is presented as an array of dimensions $H\times W$, where $H$ represents the height of the image and $W$ denotes its width, with the image possessing a single grayscale channel. This image is divided into $N$ independent patches based on the patch size $P$, where $N=(H\times W)/P^2$, and each patch forms a sequence $X_p=\{x^1,…,x^N\}$. The Generator randomly selects some of these patches to mask. Particularly for ECG data, which consists of specific combinations of waveforms, an approach that effectively masks the waveforms of ECG data is adopted instead of the conventional random patch sampling method. As a sampling strategy, the set of patch indices $P_m$ is randomly selected through a normal distribution. Considering the characteristic of ECG image data, where waveforms are distributed in the middle part, we define the set of column indices $C$ for the selected patches, which in turn defines the patches to be masked as set $M$. Subsequently, the whole image patches $X$ are divided into masked patches $X_m$ and unmasked patches $X_v$ for processing. The masking process can be formulated as follows:
\begin{equation}
P_m \sim Normal(1,N)
\end{equation}
\begin{equation}
C=\{c_i \mid (r_i,c_i )\in P_m \}
\end{equation}
\begin{equation}
M= \cup_{c\in C} \mid \{(r,c) \mid r \in \{ 1,2,…,R \},c \in C\}
\end{equation}
\begin{equation}
X_m= \{x^k \mid x^k \in M \},X_v = \{ x^k \mid x^k \notin M \}
\end{equation}

\subsubsection{Generator.} The generator follows the architecture of the MAE, which is structured to learn to reconstruct images through an encoder and decoder with input data. During the learning process, except for the masked patches $X_m$ created using the Wave Masking Strategy, the remaining visible patches $X_v$ are fed into the encoder. This encoder possesses a Transformer Encoder structure based on self-attention, generating a contextualized representation vector $R_v$ for the input data. Subsequently, the decoder receives this vector $R_v$, the masked patches $X_m$, and Positional Encoding (PE), to predict the masked patches and thereby perform the reconstruction of the entire image.

\begin{equation}
R_v=f_e (X_v)
\end{equation}
\begin{equation}
\hat{X}_m=f_d (R_v,X_m,PE)
\end{equation}
$f_e(\cdot)$ and $f_d(\cdot)$ represent the Encoder and Decoder of the Generator, respectively.
\subsubsection{Discriminator.} In the traditional structure of GANs, the discriminator is trained for the classification task of distinguishing between real images and those reconstructed by the generator. However, in anomaly detection tasks, the role of the discriminator shifts to encouraging the generator to produce images that closely resemble normal data. Through this process, the discriminator indirectly trains the generator by minimizing the difference between the reconstructed images and the real images. Notably, this model adopts the approach of training both the generator and discriminator using only normal data. As a result, the generator learns the distribution of normal data, enabling it to produce reconstructions that resemble normal data, regardless of which masked patches are input. This training method ultimately maximizes the difference between normal and abnormal data, facilitating effective anomaly detection.

\subsubsection{Training objective.} The primary training objective of our proposed model is to ensure that the generator accurately reconstructs images for normal data. Conversely, for abnormal samples, the reconstruction process is made challenging, enabling the model to classify them as such. Through this approach, the model undergoes training solely on normal data and is expected to struggle with accurate reconstruction for abnormal data due to inadequate learning of the masked patches and representation vectors. This will aid the model in identifying abnormal data. 

\paragraph{Masked Autoencoder Loss.}
The MAE that comprises the encoder and decoder of the Generator is optimized for masked patches through a mean-squared error reconstruction loss. The formula is as follows:
\begin{equation}
\mathcal{L}_{mae}(X,M,\theta_{mae})=\sum_{k\in{M}} \Vert \hat{x}^k-x^k \Vert _2^2
\end{equation}
$\theta_{mae}$ represents the parameters of the MAE model. $L_{mae}$ evaluates how similar the reconstructed image is to the original image by calculating the difference between the reconstructed patches and the original patches.

\paragraph{Adversarial Loss.}
To enhance the reconstruction capability for normal data, adversarial loss is utilized. This loss is calculated with the goal of enabling the Discriminator to classify between real patches and reconstructed patches
\begin{equation}
\mathcal{L}_{adv}(X,\theta_{mae}) = \log{D({X_v})} + \log{(1-D(\hat{X}_m))}
\end{equation}

\paragraph{Contextual Loss.}
Although the previously defined $L_{mae}$ loss function and  $L_{adv}$ loss function train the model to generate realistic data, they do not ensure that the model learns the contextual information of the input data. Learning contextual information allows the model to recognize more complex patterns and make sophisticated predictions based on these patterns. To explicitly learn contextual information, we apply L1 normalization to the input $X$ and the output $\tilde{X}$.
\begin{equation}
\mathcal{L}_{con} (X,\theta_{mae})=\left\vert X- \tilde{X} \right\vert_1
\end{equation}
Ultimately, the total training loss is determined by the weighted sum of the aforementioned losses. 
\begin{equation}
\mathcal{L}_{total} = \mathcal{L}_{mae}+ \gamma_{adv} \mathcal{L}_{adv} + \gamma_{con} \mathcal{L}_{con}
\end{equation}
$\gamma_{adv}$ and $\gamma_{con}$ are weighting parameters for the numerical stability of each loss.

\section{Experimental Results}
\subsection{Datasets}

This research employed the publicly accessible \href{https://physionet.org/content/mitdb/1.0.0/}{MIT-BIH Arrhythmia Database} for the purpose of arrhythmia detection and assessment [12]. The dataset encompasses 48 ambulatory ECG recordings sourced from 47 individuals investigated at Boston’s Beth Israel hospital arrhythmia laboratory between 1975 and 1979. Among these recordings, 23 were randomly sampled from 4000 ECGs collected from a blend group of inpatients (60\%) and out-patients (40\%) populations, while the remaining 25 included infrequent yet clinically significant arrhythmias. Each recording is dual channel, with the upper channel affixed to limb lead II (MLII/Left leg) and the lower one to the V1(V2, V4, V5) locations. Digitized at a frequency of 360 samples per second per channel, the recordings have a resolution of 11 bits and a range of 10mV, filtered from 0.1 to 100 Hz. Annotation was independently performed by two cardiac specialists. 
\\ To obtain signals at the beat level, ECG recordings with a duration of approximately 30 minutes were segmented. Segmentation was performed within the range of 0.3 to 0.4 seconds based on the R-peak, encompassing both P and T waves. Approximately 2,340 beats were generated per record. These beat-level signals were then extracted into images and resized to 128x128 after conversion to grayscale. Subsequently, following the guidelines of the Association for the Advancement of Medical Instrumentation (AAMI), 15 classes were reorganized into 5 classes: Normal(N), Supraventricular(S), Ventricular(V), Fusion(F), and Unknown(Q), as shown in Table 1 [13]. Additionally, we removed five records (102, 104, 107, 217, and 218) containing signals of poor quality, resulting in the utilization of 43 records. Consequently, a total of 100,717 data were generated, comprising 89,111 training images and 11,606 testing images. The training dataset consisted solely of N beats, while the testing dataset comprised N (n = 1,000), S (n = 2,781), V (n = 7,008), F (n = 802), and Q beats (n = 15).

\begin{table}
\centering
\caption{Reorganization of MIT-BIH heartbeat classes into AAMI classes.}\label{tab1}
\begin{tabular}{p{7cm}|p{4cm}}
\hline
MIT-BIH beat class &  AAMI beat class \\
\hline
Normal (N) &  \multirow{5}{*}{Normal (N)} \\
Left Bundle Branch Block (L) &   \\
Right Bundle Branch Block (R) &  \\
Atrial Escape (e) &  \\
Nodal (junctional) escape (j) &  \\
\hline
Atrial Premature (A) & \multirow{4}{*}{Supraventricular (S)} \\
Aberrated Atrial Premature (a) & \\
Nodal (junctional) Premature (J) & \\
Supraventricular Premature (S) & \\
\hline 
Premature Ventricular Contraction (V) & \multirow{2}{*}{Ventricular (V)} \\
Ventricular Escape (E) & \\
\hline 
Fusion of Ventricular and Normal (F) & Fusion (F) \\
\hline 
Paced (/) & \multirow{4}{*}{Unknown (Q)} \\
Fusion of Paced and Normal (f) & \\
Unclassifiable (Q) & \\
\hline 
\end{tabular}
\end{table}

\subsection{Experimental Detail}
In this study, we evaluate the performance of the EB-GAME model by training it from scratch, end-to-end, without any pre-training. The input images are divided into 64 patches, each of a size of 8x8. Additionally, a mask ratio of 0.3 was used for masking the patch images. To validate the effectiveness of the model, the generator follows the ViT-B configuration and hyperparameters applied in the standard MAE. During the training process, a dynamic token masking technique is employed, where the positions to be masked are determined during the learning phase. For model optimization, the AdamW optimizer is used, and the learning rate adjustment is achieved through a Warm-up cosine annealing scheduler. All these processes are implemented using PyTorch, with an NVIDIA RTX A4000 32GB GPU serving as the computational resource. The evaluation metrics used for validating the model include Accuracy, AUROC, Sensitivity, Specificity, and F1-Score.

\subsection{Results}
We conducted experiments for a quantitative comparison with several previous ex-periments that performed anomaly detection using the same dataset. Looking at the results of our proposed model first, it showed an AUROC of 0.97, Accuracy of 0.97, Sensitivity of 0.95, Specificity of 0.98, and also demonstrated a high F-score of 0.95. Table 2 presents the numerical comparison results with previous experiments. Upon examining the comparison results with existing GAN-based anomaly detection models, the highest performing ECG-ADGAN demonstrated an accuracy score of 0.95, an AUC score of 0.95, and an F1-score of 0.94. However, our model exhibited higher results in all evaluation metrics.

\begin{table}
\centering
\caption{Quantitative comparison results for performance evaluation.}\label{tab1}
\begin{tabular}{p{3cm}|p{2cm}|p{2cm}|p{2cm}}
\hline
Model &  Accuracy & AUROC & F1-score \\
\hline\Xhline{3\arrayrulewidth}
AnoGAN &  0.92 & 0.89 & 0.89 \\
GANomaly & 0.93 & 0.93 & 0.90 \\
BeatGAN & 0.94 & 0.94 & 0.92 \\
ECG-ADGAN & 0.95 & 0.95 & 0.94 \\
\textbf{Ours} & \textbf{0.97} & \textbf{0.97} & \textbf{0.95} \\
\hline
\end{tabular}
\end{table}

\section{Conclusion}
In this paper, we have proposed a GAN-based masked autoencoder framework for ECG heartbeat anomaly detection. Our main contribution is the introduction of a GAN-based anomaly detection framework that utilizes an ECG wave masking strategy and a MAE architecture. The framework's generator randomly masks ECG waves and predicts the masked areas. Meanwhile, the discriminator is primarily used to assist the training of the generator by comparing the characteristics of the input and reconstructed images. Our EB-GAME model demonstrated the highest performance on the dataset with an AUC score of 0.97. We plan to conduct further experiments to validate the model's general performance not only on various ECG datasets but also in signal anomaly detection tasks, aiming for a broader application scope.

%
%
%

\begin{thebibliography}{8}
\bibitem{ref_article1}
Hannun, AY., Rajpurkar, P., Haghpanahi, M., Tison, GH, Bourn, C., Turakhia, MP., Ng, A.Y.: Cardiologist-level arrhythmia detection and classification in ambulatory electrocardiograms using a deep neural network. Nature medicine 25, 65-69 (2019).

\bibitem{ref_article1}
Schull, MJ., Vermeulen, MJ., Stukel, TA.: The risk of missed diagnosis of acute myocardial infarction associated with emergency department volume. Annals of emergency medicine 48, 647-655 (2006).

\bibitem{ref_article1}
Naseer, S., Saleem, Y., Khalid, S., Bashir, MK., Han, J., Iqbal, MM., Han, K.: Enhanced network anomaly detection based on deep neural networks. IEEE access 6, 48231-48246 (2018).

\bibitem{ref_article1}
Liu, H., Zhao, Z., She, Q.: Self-supervised ECG pre-training. Biomedical Signal Processing and Control 70, 103010 (2021). 

\bibitem{ref_article1}
Shin, D-H., Park, RC., Chung, K.: Decision boundary-based anomaly detection model using improved AnoGAN from ECG data. IEEE Access 8, 108664-108674 (2020).

\bibitem{ref_article1}
Zhou, B., Liu, S., Hooi, B., Cheng, X., Ye, J.: Beatgan: Anomalous rhythm detection using adversarially generated time series. in IJCAI, Vol. 2019 4433-4439 (2019).

\bibitem{ref_article1}
Schlegl, T., Seeböck, P., Waldstein, SM., Schmidt-Erfurth, U., Langs, G.: Unsupervised anomaly detection with generative adversarial networks to guide marker discovery. International conference on information processing in medical imaging: Springer, 146-57 (2017).

\bibitem{ref_article1}
Akcay, S., Atapour-Abarghouei, A., Breckon, TP.: Ganomaly: Semi-supervised anomaly detection via adversarial training.  Computer Vision–ACCV 2018: 14th Asian Conference on Computer Vision, Perth, Australia, December 2–6, 2018, Revised Selected Papers, Part III 14: Springer, pp. 622-37 (2019).

\bibitem{ref_article1}
Brophy, E., De Vos, M., Boylan, G., Ward, T.: Multivariate generative adversarial networks and their loss functions for synthesis of multichannel ecgs. Ieee Access 9, 158936-158945 (2021).

\bibitem{ref_article1}
Qin, J., Gao, F., Wang, Z., Wong, DC., Zhao, Z., Relton, SD., Fang, H.: A novel temporal generative adversarial network for electrocardiography anomaly detection. Artificial Intelligence in Medicine 136, 102489 (2023).

\bibitem{ref_article1}
He, K., Chen, X., Xie, S., Li, Y., Dollár, P., Girshick, R.: Masked autoencoders are scalable vision learners.  in Proceedings of the IEEE/CVF conference on computer vision and pattern recognition. 16000-16009 (2022).

\bibitem{ref_article1}
Moody, GB., Mark, RG.: The impact of the MIT-BIH arrhythmia database. IEEE engineering in medicine and biology magazine 20, 45-50 (2001).

\bibitem{ref_article1}
Guan, J., Wang, W., Feng, P., Wang, X., Wang, W.: Low-dimensional denoising em-bed-ding transformer for ECG classification.  ICASSP 2021-2021 IEEE International Conference on Acoustics, Speech and Signal Processing (ICASSP), 1285-1289 (2021)

\end{thebibliography}
%

\end{document}